\documentclass[final,5p,times,authoryear]{elsarticle}
\usepackage{amsmath,amssymb,amsfonts}
\usepackage{listings}
\usepackage{xcolor}
\usepackage{graphicx}
\usepackage{textcomp}
\PassOptionsToPackage{hyphens}{url}
\usepackage{xurl}
\usepackage{booktabs}
\usepackage{array}
\usepackage{tabularx}
\usepackage[hidelinks,breaklinks=true]{hyperref}
\usepackage{algorithm}
\usepackage{algorithmic}
\usepackage{float}
\usepackage{subcaption}
\usepackage[most]{tcolorbox}



\biboptions{authoryear,round}

\journal{Computers \& Security}

\begin{document}

\begin{frontmatter}

\title{IDORacle: Template-Guided SQL-Sink Mediation for Object-Level Authorization in Java Applications}

\author[inst1]{Yuewantong Song\fnref{equal}}
\ead{sywt149@gs.zzu.edu.cn}

\author[inst2,inst3]{Guanhang Shi\fnref{equal}}
\ead{guanhang@bu.edu}

\author[inst2,inst3]{Yin Cai}
\ead{ycai25@m.fudan.edu.cn}

\author[inst2,inst3]{Changhui Wang}
\ead{changhuiwang25@m.fudan.edu.cn}

\author[inst2,inst3]{Jin Wei}
\ead{j_wei@fudan.edu.cn}

\author[inst3]{Ping Chen\corref{cor1}}
\ead{pchen@fudan.edu.cn}

\author[inst4]{Lei Shi}
\ead{shilei@zzu.edu.cn}

\author[inst1,inst3]{Jiangxing Wu}
\ead{ndscwjx@126.com}

\cortext[cor1]{Corresponding author.}
\fntext[equal]{Yuewantong Song and Guanhang Shi contributed equally to this work.}

\affiliation[inst1]{
organization={School of Computer Science and Artificial Intelligence, Zhengzhou University},
city={Zhengzhou},
country={China}
}

\affiliation[inst2]{
organization={School of Computer Science, Fudan University},
city={Shanghai},
country={China}
}

\affiliation[inst3]{
organization={Institute of Big Data, Fudan University},
city={Shanghai},
country={China}
}

\affiliation[inst4]{
organization={School of Cyber Science and Engineering, Zhengzhou University},
city={Zhengzhou},
country={China}
}

\begin{abstract}
Insecure Direct Object Reference (IDOR), frequently modeled as Broken Object-Level Authorization (BOLA), remains prevalent in Java database applications. The root cause is an architectural disconnect between identity and authorization checks enforced at the Controller or Service layer and the underlying SQL execution layer, which performs object-level operations based solely on resource identifiers. Existing research predominantly addresses detection of such vulnerabilities in source code, leaving a gap in low-intrusion runtime defense solutions for legacy Java-SQL applications.
This paper presents IDORacle, a template-guided SQL-sink interception and rewriting framework designed to mitigate horizontal privilege escalation at runtime. IDORacle propagates authenticated identity contexts across HTTP requests, asynchronous tasks, and data-access boundaries via a server-side trace identifier. At the MyBatis/JDBC boundary, it extracts SQL templates and computes dual fingerprints, performing a one-time template-level analysis to generate reusable mediation plans.
During instance execution, the framework combines runtime subject context, AST structures, table metadata, and cached authorization proofs to permit, rewrite, or block each SQL operation. IDORacle supports direct ownership predicate injection, join-derived ownership guards, probe-based authorization for group-owned resources, role-sensitive state-transition checks, and sensitive-column mediation. This guard taxonomy avoids the unsound assumption that all authorization failures reduce to a single row predicate.
To contain overhead in production environments, IDORacle architecturally decouples template-level analysis from instance-level caching. A dedicated Java-SQL benchmark suite is constructed to evaluate the framework against diverse IDOR scenarios grounded in real-world CVE reports. Experimental results demonstrate that IDORacle prevents horizontal authorization violations with a worst-case guard latency of 0.17\,ms. Its redundancy-aware optimization reduces average per-instance overhead by over 90\% (to 0.017\,ms) for hot SQL templates, bridging the gap between detection-phase BOLA analysis and practical, low-overhead runtime enforcement.
\end{abstract}


\end{frontmatter}

\section{Introduction}
\label{sec:introduction}
Broken access control remains one of the most persistent classes of web-application
security failures. In the OWASP Top 10:2025, Broken Access Control retains its
ranking as the leading application-security risk
\citep{owasp2021bac,owasp2025bac,cwe284}. In the OWASP API Security Top 10 2023, Broken Object Level Authorization (BOLA) is listed as API1:2023, reflecting the broad attack surface created when APIs expose object identifiers through URLs, request bodies, and API parameters without enforcing object-level authorization checks \citep{owasp2023bola}.
Insecure
Direct Object Reference (IDOR) is a concrete instantiation of this problem: an
authenticated user substitutes an object identifier---such as an order ID, job ID,
tenant ID, or user ID---and the server omits verification that the referenced
object falls within the user's authorized scope
\citep{cwe639,portswiggerAccessControl,portswiggerIDOR}.

Unlike injection vulnerabilities, IDOR is a semantic vulnerability. An HTTP
request may be syntactically valid, the SQL statement may be well-formed, and the
authenticated user may be permitted to invoke the endpoint, yet the concrete
database object accessed by the backend may belong to another user, tenant, or
group. Correctness therefore depends on application-specific authorization
invariants rather than on input syntax
\citep{felmetsger2010logic,pellegrino2014logicflaws,wang2026anota}. Object-level
authorization cannot be addressed as a pure input-sanitization problem; it must be
enforced at the layer where subjects, objects, and backend operations converge.

A substantial body of research has investigated static analysis, directed fuzzing,
and runtime observation to detect BOLA and missing owner-check vulnerabilities in
web applications \citep{huang2024bolaray,liu2025mocguard,liu2025bacscan,dharmaadi2025bacfuzz,zhang2025uabscan}.
These detection-oriented approaches effectively identify whether a vulnerable
endpoint exists, yet they operate within a testing-phase paradigm. In enterprise
environments, legacy codebases frequently contain large numbers of endpoints in
which authorization logic is interleaved with business SQL. Even when a scanner
successfully flags a missing ownership check, retrofitting authorization into
monolithic legacy services is error-prone and time-consuming. A deployable,
low-intrusion runtime prevention mechanism is therefore needed as a complementary
last line of defense.

A natural enforcement point is the database access boundary. Prior work has
explored fine-grained access control through query rewriting, authorization views,
row-level security, and middleware mediation
\citep{rizvi2004nontruman,chaudhuri2007predicatedgrants,mehta2017qapla,eykholt2017authorizedupdates,postgresqlRLS,sqlserverRLS,zhang2022blockaid}.
These mechanisms establish that SQL-level mediation is both security-relevant and
practical. However, applying them directly to Java web stacks exposes a critical
semantic gap: legacy Java applications enforce authentication and role checks in
filters, annotations, controllers, or service methods, while MyBatis or JDBC
mappers execute SQL statements using only resource identifiers. The authorized
subject identity is therefore separated from the SQL sink that commits the
database operation.

This paper addresses this deployment gap by studying runtime prevention of
horizontal Java-SQL authorization violations, where an authenticated
low-privilege user attempts to read, update, delete, or trigger state changes on
objects outside her authorized scope---the core semantics of IDOR/BOLA
\citep{owasp2023bfla,owasp2023bopla}. The primary goal is not to replace
framework-level authentication, RBAC, or ABAC \citep{sandhu1996rbac,hu2014abac},
but to add a last-mile SQL-sink reference monitor that complements existing checks
by requiring sufficient object-level evidence before protected SQL instances reach
the database.

Realizing this goal surfaces four challenges. First, authorization evidence is
fragmented: ownership semantics vary widely across tables, ranging from direct
ownership columns (e.g., \texttt{user\_id}) to derivation via multi-hop joins,
tenant groupings, or dynamic states. Second, SQL-sink enforcement must be
semantics-aware; uniformly appending \texttt{AND user\_id = ?} would disrupt
dictionary tables, administrator operations, and legitimate cross-user workflows.
Third, runtime enforcement must be efficient for production services, where SQL
statements execute inside loops, scheduled jobs, or asynchronous tasks, making
per-instance full parsing prohibitively costly. Fourth, credible evaluation
requires paired benign and adversarial traces covering diverse ownership patterns,
subquery-shaped bypasses, and performance stress cases.

To address these challenges, we present \textbf{IDORacle}, a template-guided
runtime prevention framework for Java-SQL IDOR vulnerabilities. IDORacle places a
lightweight mediation layer at MyBatis/JDBC-style SQL sinks. At application entry,
it propagates the authenticated subject context via a server-side trace identifier.
At the data-access boundary, it separates \emph{template planning} from
\emph{instance mediation}: a SQL template is parsed, normalized, and compiled into
a reusable mediation plan once, while per-execution work is limited to context
binding, resource-key extraction, proof lookup, and guard evaluation. Guard types
cover direct ownership predicates, join-derived ownership, group/tenant membership
probes, role-sensitive state transitions, and sensitive-column policies, avoiding
the unsound assumption that all authorization failures reduce to a single row
predicate.

This paper makes the following contributions:

\begin{itemize}
  \item \textbf{Template-guided SQL-sink reference monitor.} 
  IDORacle introduces a low-intrusion reference monitor at the database access boundary that binds trusted subject context to SQL executions. By separating template-level planning from instance-level mediation, it reuses authorization plans across normalized SQL fingerprints and enforces checks on runtime context and policy metadata for each execution.

  \item \textbf{Heterogeneous guard model for object-level authorization.} 
  IDORacle defines a guard taxonomy covering direct ownership, derived ownership, group/tenant membership, role/state policy, and sensitive-column policy. This model supports diverse authorization semantics within a unified declarative configuration, complementing existing controller- and service-layer checks.

  \item \textbf{Benchmark design and empirical evaluation.} 
  We construct a dedicated Java-SQL IDOR benchmark grounded in real-world CVE reports to evaluate the framework across diverse ownership patterns. Results demonstrate that IDORacle prevents all tested horizontal privilege escalation attempts with zero false positives. The redundancy-aware caching mechanism reduces average overhead by over 90\% (to 0.017\,ms) for hot SQL templates, with a worst-case latency of 0.17\,ms.
\end{itemize}

\section{Background}
\label{sec:background}

\subsection{Scoping the Defense: BOLA, BFLA, and BOPLA}
\label{subsec:ac_taxonomy}

Broken access control is a broad category. For this paper, three subcategories
are especially relevant. \emph{Broken Object-Level Authorization} concerns
whether the authenticated subject is allowed to access the concrete object
referenced by a request. BOLA is the API-security formulation of IDOR: a user
may be allowed to call an endpoint, but not allowed to access the object
identified by the supplied parameter \citep{owasp2023bola,cwe639}. This is
usually a horizontal authorization problem when users with similar roles access
one another's objects.

\emph{Broken Function-Level Authorization} concerns whether a subject is
allowed to invoke a function or endpoint at all. BFLA is commonly associated
with role hierarchies, administrator functions, and missing function-level
checks \citep{owasp2023bfla}. A user invoking an administrator-only endpoint is
therefore a function-level or vertical authorization problem, even if the SQL
statement later touches ordinary rows. \emph{Broken Object Property-Level
Authorization} concerns whether a subject is allowed to access or modify
particular properties of an otherwise accessible object, such as passwords,
tokens, phone numbers, addresses, or internal status fields \citep{owasp2023bopla}.

This distinction matters for system design. A SQL-sink defense is well-positioned
for BOLA/IDOR because the SQL statement exposes the target table, operation,
predicate, and bound object identifiers immediately before database execution.
By contrast, a pure SQL-layer mechanism cannot fully infer whether a route is
administrator-only without additional route-permission evidence. Similarly,
property-level authorization may require column policies rather than row
ownership predicates. IDORacle therefore treats horizontal object-level
authorization as its primary target and handles function-level and
property-level cases only when sufficient evidence can be carried to the SQL
boundary.

\subsection{The Architectural Root Cause of Java-SQL IDOR}
\label{subsec:java_idor}

Java web applications commonly distribute authorization logic across multiple
layers. Authentication may be enforced in a servlet filter or gateway. Coarse
permission checks may be implemented using annotations, route rules, Shiro,
Spring Security, or custom interceptors. Business-specific ownership checks may
appear in service methods. Finally, MyBatis XML mappers, annotation-based SQL,
or JDBC calls execute concrete SQL statements. This separation is useful for
engineering modularity, but it also introduces a semantic gap: the layer that knows the current subject may not be the layer that observes the concrete database object.

To illustrate why a single hard-coded fix is insufficient, consider the two 
vulnerable scenarios shown below:

\begin{lstlisting}[basicstyle=\ttfamily\footnotesize, escapechar=|, frame=single, backgroundcolor=\color{gray!5}]
|\textcolor{teal}{// [Scenario A] Direct Ownership (Easy)}|
Controller: requireLogin()
Mapper: |\textcolor{blue}{SELECT}| * |\textcolor{blue}{FROM}| t_order 
        |\textcolor{blue}{WHERE}| id = ?
|\textcolor{red}{// Missing: AND user\_id = \{curr\_user\}}|

|\textcolor{teal}{// [Scenario B] Join-Derived (Complex)}|
Controller: requireLogin()
Mapper: |\textcolor{blue}{SELECT}| * |\textcolor{blue}{FROM}| t_order_item 
        |\textcolor{blue}{WHERE}| item_id = ?
|\textcolor{red}{// Missing: Requires JOIN t\_order ON ...}|
|\textcolor{red}{//        AND t\_order.user\_id = \{curr\_user\}}|
\end{lstlisting}

In both scenarios, the endpoint authenticates the user and the SQL selects a valid
row, but no component verifies that the referenced resource belongs to the current
user. The bug is not a missing login check; it is a missing object-level invariant.
A further complication arises from the widespread use of database connection pools
(e.g., HikariCP, Druid) in Java enterprise applications. Because all application
users share the same database credentials, the underlying DBMS has no visibility
into the end-user identity associated with each query, making native enforcement
of user-specific constraints infeasible without complex session-variable
manipulations in legacy configurations.

\begin{figure*}[t]
    \centering
    \includegraphics[width=0.98\textwidth]{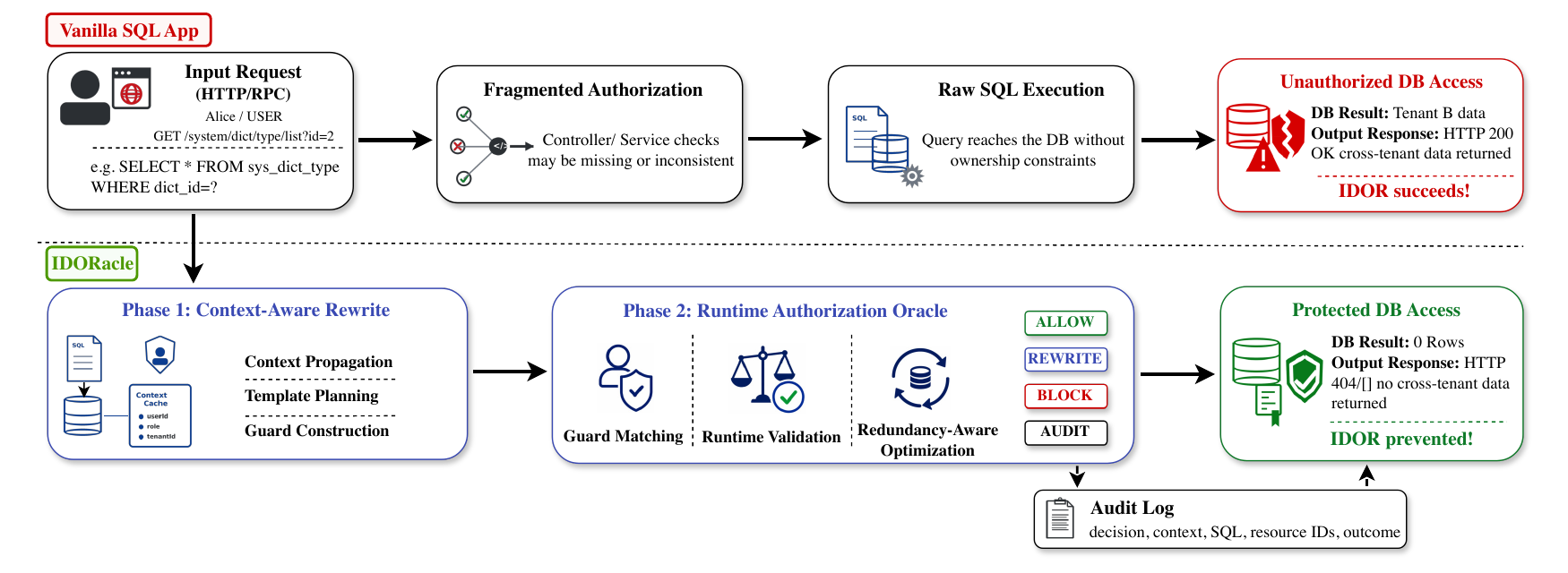}
    \caption{Overview of IDORacle comparing the Vanilla SQL App execution path with the template guided mediation path.}
    \label{fig:idoracle-overview}
\end{figure*}

The persistence of IDOR directly follows from this heterogeneity of ownership 
semantics. While Scenario A can be fixed by injecting a simple \texttt{user\_id} 
predicate, Scenario B derives ownership indirectly. Furthermore, some resources 
may belong to a tenant rather than a user, and administrative endpoints may bypass 
these checks entirely. These cases demonstrate that horizontal authorization cannot 
be handled by a naive, one-size-fits-all row predicate.

\subsection{From Detection to Runtime Enforcement}
\label{subsec:detection_background}

Existing work on BOLA and missing owner-check vulnerabilities spans static analysis, directed fuzzing, and differential testing~\citep{huang2024bolaray,liu2025mocguard,liu2025bacscan,zhang2025uabscan}. These approaches effectively identify whether a vulnerable endpoint exists, but they operate within a testing-phase paradigm: scanners flag missing checks before deployment, leaving legacy production codebases unprotected. Even when a scanner successfully identifies a flaw, retrofitting authorization constraints across a large monolithic service is error-prone and time-consuming.

Database-level mechanisms such as row-level security (RLS) and query-rewriting frameworks~\citep{rizvi2004nontruman,postgresqlRLS,mehta2017qapla} establish that SQL-layer mediation is feasible, but they assume a one-to-one mapping between database users and application users---an assumption broken by the connection-pool architecture described in Section~\ref{subsec:java_idor}. Furthermore, they do not bind Java runtime subject context to individual SQL instances across asynchronous boundaries, and they cannot express the heterogeneous ownership policies (direct, derived, group, and role-sensitive) found in real Java applications.

These two gaps together define the design space for IDORacle: a runtime enforcement layer that mediates SQL executions against authenticated subject context, operates with no source-code changes to legacy services, and amortizes the cost of policy evaluation through \emph{ahead-of-time template planning} rather than per-query dynamic rewriting. Detailed comparison with related detection and enforcement systems is provided in Section~\ref{sec:related-work}.

\section{Threat Model}
\label{sec:threat_model}

\subsection{Attacker Capabilities and Objectives}
The adversary is an authenticated low privilege user of a Java database application. The adversary can send ordinary HTTP or RPC requests and freely modify request controlled identifiers, including path variables, query parameters, request body fields, and RPC arguments. These identifiers may later be bound to SQL statements as business keys. The adversary aims to read, update, delete, or trigger state changes on objects outside their authorized scope. The adversary may exploit missing owner checks, incomplete service validation, controller permission omissions, subquery shaped SQL paths, or inconsistent treatment of group and tenant resources. The adversary does not have database credentials, cannot directly modify the policy metadata store, and cannot forge authenticated identities or server trace bindings.

\subsection{Defender Capabilities and Assumptions}
The defender controls the Java application deployment, the SQL interception component, and the authorization metadata used by IDORacle. The defender can recover authenticated subject context from the existing security framework, register table and column policies, and deploy the system in shadow or enforce mode. The defender does not assume that all controllers, services, or mapper SQL statements are correctly protected. The trusted computing base consists of the entry context extractor, the context store, the SQL hook, the metadata store, the mediator, as well as the underlying JVM environment and the database engine. These components are assumed to preserve the integrity of the interception agent and SQL execution. IDORacle assumes that the upstream identity propagation infrastructure is trustworthy and uncompromised.
\section{Methodology}
\label{sec:methodology}

\subsection{Overview}
\label{subsec:method_overview}

As illustrated in Figure~\ref{fig:idoracle-overview}, the architectural vulnerability emerges when an authenticated user, such as Alice, submits a data retrieval request triggering an underlying database query. Within the Vanilla SQL App execution path, fragmented authorization allows this request to bypass incomplete controller or service checks. Consequently, the raw SQL execution proceeds directly to the database without enforcing ownership constraints. This structural flaw leads to unauthorized database access, ultimately yielding an HTTP 200 OK response that exposes cross tenant data and signifies a successful IDOR attack.

For the identical input request, the IDORacle execution path mitigates the vulnerability through a two phase template guided mediation pipeline. Phase 1 executes a context aware rewrite by binding the TraceID security context of the user and intercepting the query at the DAO and JDBC boundary to generate a secured SQL candidate. Phase 2 then engages the runtime authorization oracle to evaluate this candidate via guard matching, runtime validation, and redundancy aware optimization. The oracle outputs a definitive action, selecting among allow, rewrite, block, or audit decisions, and records the outcome in an audit log. By enforcing strict ownership boundaries prior to execution, this pipeline guarantees protected database access. The database evaluates the secured query to yield zero rows, compelling the application to return a safe HTTP 404 or empty list response, thereby preventing cross tenant data leakage and neutralizing the IDOR attempt.

The IDORacle design does not replace existing framework level authentication or role authorization. Instead, it treats these mechanisms as evidence sources and revalidates the concrete object level operation at the SQL sink. This positioning is particularly relevant for legacy Java applications, where the controller may encode route permissions, the service may apply partial business checks, and the SQL mapper ultimately executes the database operation.

\subsection{Phase 1: Context-Aware Rewrite}
\label{subsec:phase1}

Phase~1 establishes the trusted identity anchor for each SQL execution and
compiles a reusable mediation plan before any concrete parameter values are
bound. It comprises three sequential steps that correspond to the three
operational lines shown in Figure~\ref{fig:idoracle-overview}: binding the
TraceID security context, intercepting the SQL at the Data Access Object
(DAO)/JDBC boundary, and generating the secured SQL candidate that is
forwarded to Phase~2 for evaluation.

\paragraph{Context propagation}
A SQL instance can be mediated only if it is bound to the authenticated subject
that triggered it. IDORacle therefore propagates a server-side identity record
rather than trusting user-controlled request parameters. At the application
entry point, the system extracts the authenticated principal and stores a
compact identity record (Eq.~\eqref{eq:context}):
\begin{equation}
\begin{aligned}
\texttt{ctx:\{traceId\}} \mapsto
\{&\texttt{userId},\ \texttt{role},\ \texttt{tenantId},\\
  &\texttt{sessionAttr},\ \texttt{policyVersion}\}.
\end{aligned}
\label{eq:context}
\end{equation}
Only the identity record is propagated through Mapped Diagnostic Context (MDC),
Remote Procedure Call (RPC) attachments, message headers, or scheduled-task
wrappers to reliably bridge the semantic gap across thread boundaries; this
follows the general practice of carrying opaque execution context rather than
trusting user-supplied identity fields~\citep{w3cTraceContext}. The full
identity record remains in the trusted server-side store. At the SQL sink, the
mediator first checks a short-lived local cache and then falls back to the
shared context store. Missing or expired context is treated as a fail-closed
condition unless the callsite is explicitly registered as a system task. This
rule prevents silently allowing attacker-triggered SQL statements when context
propagation fails.

\paragraph{Template planning}
IDORacle separates SQL-template analysis from per-instance mediation. The SQL
hook intercepts a parameterized SQL template at the DAO/JDBC boundary before
concrete values are bound and constructs a template fingerprint $h_T$
(Eq.~\eqref{eq:template_hash}):
\begin{equation}
\begin{split}
h_T = H(&\textit{mapperId}, \textit{normalizedAst}, \textit{commandType}, \\
       &\textit{parameterMappings}, \textit{callsiteHash}, \textit{policyVersion}).
\end{split}
\label{eq:template_hash}
\end{equation}
Including the mapper identifier, parameter mapping, callsite hash, and policy
version prevents unrelated SQL statements with similar Abstract Syntax Tree
(AST) shapes from sharing an authorization plan. This strict fingerprinting
reduces the risk that template plans are inadvertently shared across
semantically distinct SQL statements or invalidated policy versions. If the
fingerprint is unseen, the planner normalizes the SQL, parses its AST, extracts
command type, tables, joins, subqueries, selected or updated columns, and
predicate-related parameter positions, and then binds these facts to policy
metadata. The output is a \emph{TemplatePlan} (summarized in
Table~\ref{tab:template_plan}), which is cached locally and can be replicated
to a distributed cache in multi-instance deployments.

\begin{table}[t]
\centering
\caption{Main fields in a TemplatePlan.}
\label{tab:template_plan}
\small
\begin{tabular}{p{0.28\linewidth}p{0.60\linewidth}}
\toprule
\textbf{Field} & \textbf{Description} \\
\midrule
\texttt{templateHash} & Fingerprint of the normalized SQL template and its execution context. \\
\texttt{operation} & SQL command type, such as SELECT, UPDATE, DELETE, or INSERT. \\
\texttt{tables} & Database tables touched by the SQL AST, including subqueries. \\
\texttt{columns} & Selected, updated, or predicate-related columns. \\
\texttt{riskClass} & Public table, direct ownership, derived ownership, group ownership, state transition, sensitive column, or admin-only access. \\
\texttt{guards} & Metadata-bound guards that must be evaluated for the template. \\
\texttt{resourceKeys} & Parameter positions or expressions used to derive protected resources. \\
\texttt{action} & Planned mediation action: allow, rewrite, probe, block, mask, or audit. \\
\bottomrule
\end{tabular}
\end{table}

Template planning is the principal performance optimization. Java applications
often execute a small number of SQL templates many times in loops, scheduled
jobs, or branch-heavy business logic. IDORacle therefore performs expensive
parsing, normalization, table classification, and policy binding once per
template and keeps per-execution work limited to identity record binding,
parameter extraction, proof lookup, and guard-specific mediation.

\paragraph{Guard construction}
The planner attaches guards to a SQL template according to the metadata of the
touched tables, columns, and operations. A guard describes how authorization
should be established for a protected database object. To minimize deployment
overhead, these guards and associated metadata are defined via a centralized,
declarative configuration (e.g., YAML or JSON) maintained by the defender.
This centralized approach eliminates the need to scatter redundant authorization
checks across various services, ensuring easier auditing and uniform
enforcement. Table~\ref{tab:guards} lists the guard types used by IDORacle.

\begin{table}[t]
\centering
\caption{Guard types used for Java-SQL IDOR mediation.}
\label{tab:guards}
\small
\begin{tabularx}{\linewidth}{@{}p{0.31\linewidth}X@{}}
\toprule
\textbf{Guard} & \textbf{Purpose} \\
\midrule
Direct ownership & Adds or verifies an owner predicate when the target table has an explicit owner column. \\
Derived ownership & Derives authorization through a parent table, foreign-key relation, or configured join path. \\
Group/tenant membership & Resolves group or tenant membership through a membership table or resource probe. \\
Role/state policy & Checks actor role, forbidden self-approval, or allowed state transition. \\
Sensitive-column policy & Blocks, masks, or audits unauthorized access to protected columns. \\
Public table & Allows explicitly registered dictionary or configuration tables. \\
Admin-only policy & Requires administrator or privileged-route evidence for protected operations. \\
\bottomrule
\end{tabularx}
\end{table}

A table can carry multiple guards. For example, an order table may use direct
ownership for ordinary user operations and an admin-only guard for back-office
operations. A tenant-membership table may combine group membership and
state-transition guards. A user table may combine direct ownership with
sensitive-column mediation. Table~\ref{tab:metadata_example} illustrates
representative policy metadata entries. The guards are deliberately explicit:
IDORacle does not infer ownership from arbitrary column names alone, because
such inference would create both false positives and unsafe bypasses.

\begin{table}[t]
\centering
\caption{Examples of authorization metadata.}
\label{tab:metadata_example}
\small
\begin{tabularx}{\linewidth}{@{}p{0.40\linewidth}X@{}}
\toprule
\textbf{Object} & \textbf{Policy metadata} \\
\midrule
\texttt{t\_order} & Direct owner column \texttt{user\_id}. \\
\texttt{t\_order\_item} & Derived owner via \texttt{order\_id} $\rightarrow$ \texttt{t\_order.id}. \\
\texttt{xxl\_job\_info} & Group owner via \texttt{job\_group}. \\
\texttt{sys\_user\_tenant} & Role/state guard for tenant approval. \\
\texttt{sys\_user.password} & Sensitive column blocked for non-admin users. \\
\texttt{t\_sku\_dict} & Public dictionary table. \\
\bottomrule
\end{tabularx}
\end{table}

\subsection{Phase 2: Runtime Authorization Oracle}
\label{subsec:phase2}

Phase~2 evaluates the secured SQL candidate produced by Phase~1 against the
runtime identity record and bound parameters, issuing one of four decisions:
\textsc{Allow}, \textsc{Rewrite}, \textsc{Block}, or \textsc{Audit} (shown in\
Figure~\ref{fig:idoracle-overview}). The oracle comprises three concurrent
sub-components---guard matching, runtime validation, and redundancy-aware
optimization---formalized in Algorithm~\ref{alg:mediation}.

\paragraph{Guard matching}
At runtime, the instance mediator evaluates the \emph{TemplatePlan} against
the current identity record and bound parameters. If the SQL template is
registered as a public table access, the mediator issues an \textsc{Allow}
decision immediately. If the template touches a protected table, the mediator
extracts concrete resource keys according to the plan and evaluates the
attached guards.

For direct ownership, the guard inserts a trusted owner predicate using the SQL
AST rather than string concatenation:
\begin{center}
\begin{minipage}{0.96\linewidth}
\footnotesize\ttfamily
SELECT * FROM t\_order WHERE id = ?\\
$\Rightarrow$\\
SELECT * FROM t\_order WHERE id = ? AND user\_id = ?
\end{minipage}
\end{center}
The appended parameter is bound from \texttt{ctx:\{traceId\}}, not from the
attacker-controlled request. UPDATE and DELETE statements are constrained in
the same way so that unauthorized rows are not modified.

For derived ownership, the mediator uses metadata to connect a child table to
an owner-bearing parent table. When safe SQL rewriting is possible---typically
in single-table queries or straightforward joins---it inserts an \texttt{EXISTS}
predicate:
\begin{center}
\begin{minipage}{0.96\linewidth}
\footnotesize\ttfamily
SELECT * FROM t\_order\_item i WHERE i.id = ?\\
$\Rightarrow$\\
SELECT * FROM t\_order\_item i WHERE i.id = ?\\
\quad AND EXISTS (SELECT 1 FROM t\_order o\\
\quad WHERE o.id = i.order\_id AND o.user\_id = ?)
\end{minipage}
\end{center}
However, AST rewriting can inadvertently alter query semantics. If the SQL
dialect, complex aggregations (\texttt{GROUP BY}), unions (\texttt{UNION}), or
certain Data Manipulation Language (DML) structures render rewriting unsafe,
the mediator gracefully falls back to a proof query
(\textsc{Probe-then-Allow}) under the same request trace.

For group- or tenant-owned resources, the mediator first resolves the protected
resource and then checks membership. A job operation, for example, may receive
only \texttt{job\_id}, while authorization is defined over \texttt{job\_group}.
The mediator resolves the group and verifies membership before the original SQL
is permitted.

\paragraph{Runtime validation}
Not every protected operation can be expressed as row ownership. A tenant-join
approval may require the actor to be an administrator of the tenant and not the
applicant whose request is being approved. A user-list query may be legitimate
for ordinary users but unsafe if it returns password hashes, tokens, phone
numbers, or addresses. IDORacle handles these cases with role/state and
sensitive-column guards.

A role/state guard checks the actor role, the protected resource, the old
state, and the requested transition before the DML statement is executed. When
the condition can be expressed safely in SQL, it is compiled into an
\texttt{EXISTS} predicate. Otherwise, the mediator issues a proof query and
blocks the update if the proof fails. A sensitive-column guard inspects the
SELECT projection and applies the configured action: block, rewrite to
\texttt{NULL}, mask the column, or audit. For example:
\begin{center}
\begin{minipage}{0.96\linewidth}
\footnotesize\ttfamily
SELECT user\_id, username, password, mobile FROM sys\_user\\
$\Rightarrow$\\
SELECT user\_id, username, NULL AS password,\\
\quad CONCAT('***', RIGHT(mobile, 4)) AS mobile FROM sys\_user
\end{minipage}
\end{center}
High-risk columns such as passwords and tokens are blocked by default unless
route and role evidence explicitly authorize the access.

\paragraph{Redundancy-aware optimization}
The result of each guard evaluation is cached as an authorization proof $h_P$
(Eq.~\eqref{eq:proof_hash}):
\begin{equation}
\begin{split}
h_P = H(&\textit{traceId}, \textit{subjectHash}, \textit{templateHash}, \\
       &\textit{operation}, \textit{resourceKey}, \textit{policyVersion}).
\end{split}
\label{eq:proof_hash}
\end{equation}
The authorization proof records the decision, evidence source, resolved
resource attributes, guard type, and expiration time. To balance performance
and consistency in concurrent environments, authorization proofs are assigned a
short time-to-live (TTL). Write operations trigger a best-effort cache eviction
for the affected resource keys, while policy-version changes immediately
invalidate all stale authorization proofs. This mechanism is directly reflected
in Algorithm~\ref{alg:mediation} (lines~6--8), where a cache hit
short-circuits guard evaluation entirely, eliminating redundant mediation for
hot SQL templates that recur in loops, paginated queries, or scheduled tasks.

\begin{algorithm}[H]
\caption{SQL-Sink Mediation Logic}
\label{alg:mediation}
\footnotesize
\begin{algorithmic}[1]
\REQUIRE \parbox[t]{0.72\linewidth}{SQL template $T$; bound parameters $B$; identity record $r$}
\ENSURE Mediated SQL AST, or a \textsc{Block} decision
\STATE $C \gets \textsc{LoadContext}(r)$
\IF{$C=\varnothing$ \AND $\neg\textsc{IsSystemTask}(T)$}
    \RETURN \textsc{Block} \COMMENT{fail closed}
\ENDIF
\STATE $P \gets \textsc{GetOrCompilePlan}(T)$
\IF{$P.\textit{riskClass}=\textsc{Public}$}
    \RETURN \textsc{Allow}
\ENDIF
\STATE $K \gets \textsc{ExtractResourceKey}(B,P)$
\STATE $h_P \gets H(r,C,P,K)$
\IF{$\textsc{CacheHas}(h_P)$}
    \RETURN $\textsc{CacheGet}(h_P)$
\ENDIF
\STATE $D \gets \textsc{EvaluateGuards}(P.\textit{guards}, C, K)$
\STATE $\textsc{CachePut}(h_P,D)$ \COMMENT{short TTL}
\RETURN $D$
\end{algorithmic}
\end{algorithm}

\subsection{Protected Database Access and Enforcement Outcomes}
\label{subsec:enforcement_outcomes}

The oracle decision produced by Phase~2 determines the observable outcome at
both the database layer and the application response layer, corresponding to
the Protected DB Access box in Figure~\ref{fig:idoracle-overview}. When the
identity record is bound and the \emph{TemplatePlan} carries a valid
ownership-constrained SQL candidate, the database evaluates the rewritten
statement and returns zero rows for unauthorized object references, compelling
the application to return a safe HTTP~404 or an empty list response. No
cross-tenant data is exposed, and the IDOR attempt is neutralized. Each
decision is recorded in the audit log with the identity record, SQL
fingerprint, guard type, and outcome timestamp.

\paragraph{Operational modes}
IDORacle supports two modes governing how oracle decisions are applied. In
\emph{shadow mode}, the framework records decisions and potential SQL
differences without altering application behavior; this mode is used to
validate metadata configurations, estimate false-positive rates, and identify
legitimate cross-object queries that require explicit policy exceptions prior
to production enforcement. In \emph{enforce mode}, the oracle decision is
applied before the SQL statement reaches the database. An \textsc{Allow}
decision permits the original SQL to execute unmodified. A \textsc{Rewrite}
decision substitutes the original AST with the ownership-constrained SQL
candidate, producing the zero-row database result and safe HTTP response
illustrated in Figure~\ref{fig:idoracle-overview}. A \textsc{Block} decision
prevents database execution entirely. An \textsc{Audit} decision permits
execution while recording a tamper-evident log entry.

\paragraph{Fail-closed strategy}
To ensure operational resilience, IDORacle adopts a fail-closed strategy for
all critical error conditions. Missing identity record, absent policy metadata
for protected tables, conflicting owner predicates, failed membership probes,
and unauthorized sensitive-column accesses are all blocked by default rather
than permitted. This design guarantees that uncertainty in the authorization
state is resolved conservatively, preventing attacker-triggered SQL statements
from executing silently when context propagation fails or metadata is
incomplete.

\paragraph{Explicit exception management}
Explicit allow rules are required for public dictionary tables, migration
tasks, scheduled system jobs, and administrator-only workflows. These
exceptions are maintained in declarative metadata rather than in ad hoc code
comments, ensuring that the policy configuration remains auditable and
version-controlled. Original SQL comments are normalized and are not trusted as
authorization evidence. When an audit comment is emitted, it carries only an
opaque proof identifier; the corresponding evidence is stored exclusively in
the trusted server-side channel, preventing comment-injection bypasses.
\section{Benchmark Design}
\label{sec:benchmark}

The benchmark is designed to evaluate whether SQL-level runtime mediation can
prevent IDOR vulnerabilities when object-level authorization is missing,
incomplete, or inconsistently placed in Java application code. Instead of
testing only whether an HTTP endpoint is reachable, each benchmark case follows
the complete path from request to database effect. A case therefore specifies
the authenticated principal, the attacker-controllable request field, the SQL
template issued by the application, the protected-table metadata, and the
expected database and response-level outcomes.

\subsection{Design Goals}
\label{subsec:benchmark_goals}

The benchmark has three primary goals. First, it covers common object-level
authorization failures in Java-SQL applications, including unauthorized reads,
updates, deletes, and batch operations. Second, it includes ownership patterns
that cannot be handled by a single hard-coded predicate, such as ownership
derived through a parent table, group membership, tenant membership, or a
configured resolver relation. Third, it stresses the runtime cost of mediation
by including repeated SQL templates, loop-generated mapper calls, and
branch-heavy service paths. These goals allow the benchmark to measure both the
security efficacy of IDORacle and the practical benefit of its
redundancy-aware optimization.

\subsection{Principals and Protected Objects}
\label{subsec:benchmark_principals}

The benchmark uses a compact but heterogeneous Java-SQL data model. The schemas
cover users, orders, addresses, dictionary records, notices, scheduled jobs,
job groups, job logs, tenants, and user-profile data. In total, the benchmark comprises comprehensive vulnerability scenarios synthesized from real-world CVE reports affecting open-source Java applications. The RuoYi-derived cases are grounded in public CVE and advisory records \citep{nvdCVE202528400,nvdCVE202528402,nvdCVE202528406,nvdCVE202528407,nvdCVE202528411,nvdCVE202528412,nvdCVE202528413,nvdCVE202570985,ruoyiCvePublic,ruoyiProject}, while the additional XXL-Job and BootDo cases are grounded in their public vulnerability and project sources \citep{nvdCVE202333779,xxljobProject,bootdoAdvisory2024,bootdoProject}. The protected tables are
selected to represent different authorization shapes. For example,
\texttt{sys\_dict\_type} and \texttt{sys\_notice} use direct ownership through
\texttt{create\_by}; \texttt{datax\_job\_info} uses direct ownership through a
numeric \texttt{user\_id}; \texttt{xxl\_job\_info} uses group-derived ownership
through \texttt{job\_group}; and \texttt{sys\_job\_log} uses resolver-style
ownership because a log record must be linked back to an owner-bearing job
record.

The principal set contains at least two low-privilege users from different
ownership domains and an administrator. This setup separates three behaviors:
legitimate self-access, horizontal cross-user or cross-tenant access, and
legitimate administrative access. Each protected object is associated with the
security-context field that should be used as authorization evidence, such as
\texttt{userId}, \texttt{username}, \texttt{role}, or \texttt{tenantId}. The same metadata is used by the runtime oracle during evaluation, as summarized
later in Table~\ref{tab:realworld_dataset}.

\subsection{Paired Benign and Adversarial Traces}
\label{subsec:benchmark_traces}

Each functional scenario is encoded as a pair of traces. The benign trace uses
a resource identifier that belongs to the authenticated principal and should
preserve the original application behavior. The adversarial trace keeps the
same principal and endpoint but changes the identifier, transition target, or
batch item to a resource owned by another user, group, or tenant. This paired
construction is critical to demonstrate that IDORacle mitigates the vulnerability without introducing false positives that break legitimate workflows.

The overview case follows this construction. Alice is authenticated as a normal
user and sends the following request:

\begin{lstlisting}[basicstyle=\ttfamily\footnotesize, frame=single, backgroundcolor=\color{gray!5}]
GET /system/dict/type/list?id=2
\end{lstlisting}

The application issues the SQL template:

\begin{lstlisting}[basicstyle=\ttfamily\footnotesize, frame=single, backgroundcolor=\color{gray!5}]
SELECT * FROM sys_dict_type WHERE dict_id = ?
\end{lstlisting}

In the vanilla application, the query reaches the database without an ownership
constraint and may return a dictionary record belonging to another user. Under
IDORacle, the same SQL execution is mediated at the DAO/JDBC boundary. The
secured SQL candidate adds the trusted ownership predicate:

\begin{lstlisting}[basicstyle=\ttfamily\footnotesize, frame=single, backgroundcolor=\color{gray!5}]
SELECT * FROM sys_dict_type WHERE dict_id = ? 
AND create_by = ?
\end{lstlisting}

The additional value is derived from the server-side security context rather
than from attacker-controlled request parameters. Therefore, the adversarial
trace returns zero rows or a safe response such as \texttt{HTTP 404} or an
empty list, while the benign trace remains accessible.

\subsection{Case Families}
\label{subsec:benchmark_families}

The benchmark contains four case families:

\textbf{Object-level IDOR cases.}
These cases evaluate the core threat. Horizontal read cases test whether a user
can retrieve another user's row by modifying an identifier in the request.
Horizontal update and delete cases test whether the same manipulation can
modify or remove unauthorized rows. Batch cases model service code that
iterates over a list of attacker-supplied identifiers and invokes the same
mapper method repeatedly. The expected protected behavior is either a rewritten
SQL statement whose ownership predicate reduces the result or affected rows to
zero, or a \textsc{Block} decision before database execution.

\textbf{Derived-ownership cases.}
These cases cover objects whose authorization evidence is not stored directly
in the accessed table. Join-derived cases require the mediator to constrain the
target row through a related parent table. Group- and tenant-derived cases
require the oracle to validate whether the target group or tenant is in the
current subject's permission scope. Resolver cases require an \texttt{EXISTS}
predicate or an equivalent resolver query to connect the accessed record to an
owner-bearing table. These cases prevent the benchmark from overfitting to
direct \texttt{owner = subject} predicates.

\textbf{Policy-boundary cases.}
These cases check whether the mediator distinguishes protected object
resources from legitimate public or administrative accesses. Public dictionary
or configuration tables should be allowed when they are not registered as
protected objects. Administrator-only operations should be allowed only when
the authenticated role satisfies the configured policy. Column-sensitive cases
are retained as an extension for sensitive-field protection, but they are not
the main focus of the overview because the central threat in this work is
object-level IDOR rather than general data masking.

\textbf{Optimization stress cases.}
These cases expose redundant mediation work. They include paginated list
queries that repeatedly execute the same SQL template with different offsets,
batch updates or deletes that invoke the same mapper method inside a loop, and
switch-branch service paths where different actions reach the same protected
table. These cases evaluate whether template-plan reuse, metadata reuse, and
request-local deduplication reduce repeated parsing, guard matching, and
rewrite-decision generation without changing the security outcome.

\subsection{Expected Outcomes}
\label{subsec:benchmark_outcomes}

For each case, the benchmark defines both a database-level outcome and a
response-level outcome. In the vanilla mode, an adversarial trace is expected
to reach the database without an ownership constraint and may return or modify
cross-user or cross-tenant data. In the protected mode, IDORacle produces one
of four outcomes: \textsc{Allow}, \textsc{Rewrite}, \textsc{Block}, or
\textsc{Audit}. A benign trace should be allowed or rewritten without changing
the legitimate result. An adversarial read should not return the unauthorized
row. An adversarial update or delete should either affect zero rows after
rewriting or be blocked before execution. An audit record should contain the
identity record, subject context, SQL fingerprint, protected table, decision, and final outcome.

This decoupled expected-outcome design separates security correctness from application
presentation. Different Java applications may translate an empty protected
result into \texttt{HTTP 404}, an empty list, or a generic forbidden response.
The benchmark therefore treats the absence of unauthorized data at the database level as the core
security condition and records the concrete HTTP response as supporting evidence.

\subsection{Execution Modes and Measurements}
\label{subsec:benchmark_modes}

The benchmark is executed in three modes, designed to isolate the performance impact of IDORacle's core mechanisms:

\begin{itemize}
  \item \textbf{NO\_GUARD}: SQL statements are executed as they would be in the vanilla application. This mode provides the vulnerable baseline and exposes the unauthorized database result for adversarial traces.
  \item \textbf{GUARD\_NO\_CACHE}: Every guarded SQL statement goes through the full runtime oracle path, including SQL normalization, guard matching, metadata lookup, ownership validation, and rewrite construction. This mode represents the worst-case protected execution path.
  \item \textbf{GUARD\_WITH\_CACHE}: \textbf{IDORacle} enables redundancy-aware optimization and reuses decisions for repeated normalized SQL templates and repeated resource checks. This mode represents the common hot-template path in production Java applications.
\end{itemize}

For every execution, the benchmark driver records the original SQL, the
rewritten SQL when applicable, the oracle decision, the affected rows, the
response class, the cache hit or miss status, and the processing time. These logs are used to compute security metrics---including successful attack
prevention rate and false-positive rate---as well as performance metrics,
including average overhead, maximum overhead, and cache effectiveness.

\section{Implementation}
\label{sec:implementation}

We implemented IDORacle as a low-intrusion Java-SQL runtime defense prototype.
The prototype comprises approximately 5.8K lines of self-developed Java code, excluding third-party benchmark applications and framework dependencies, organized into two major components: (1) a backend policy engine ($\sim$4.0K lines) for ownership metadata management, access-control decision-making, and SQL rewriting; and (2) a Java agent ($\sim$1.8K lines) for runtime JDBC interception, identity propagation, and transparent SQL enforcement. Table~\ref{tab:impl_modules} lists the six implementation modules. Concretely, the system exposes three deployment artifacts: a guard server, an in-process MyBatis interceptor for the evaluation testbed, and the Java agent for protecting external applications without source-code modification.

\begin{table}[t]
\centering
\caption{Implementation modules of IDORacle}
\label{tab:impl_modules}
\small
\begin{tabularx}{\linewidth}{l l >{\raggedright\arraybackslash}X}
\toprule
\textbf{Module} & \textbf{Technology} & \textbf{Responsibility} \\
\midrule
Guard Server & Spring Boot & Policy management and rewrite service \\
SQL Rewriter & JSqlParser & SQL parsing and AST rewriting \\
Java Agent & Byte Buddy & Runtime SQL interception \\
Identity Layer & Servlet/Shiro Hooks & Identity extraction and propagation \\
Rewrite Cache & ConcurrentHashMap & Rewrite decision reuse \\
Observability UI & React + Vite & Visualization and benchmark control \\
\bottomrule
\end{tabularx}
\end{table}

The guard server is implemented in Spring Boot. It maintains the runtime switch, identity record, guard decisions, evaluation logs, and table-level ownership metadata. The metadata also supports administrator bypasses for cases where cross-user access is legitimate.

For SQL interception inside the evaluation testbed, IDORacle uses a MyBatis \texttt{StatementHandler.prepare} interceptor. Before SQL reaches the database, the interceptor obtains the current identity record, parses the SQL with JSqlParser \citep{jsqlparser}, detects the guarded table, and injects an ownership predicate into \texttt{SELECT}, \texttt{UPDATE}, and \texttt{DELETE} statements.

For a direct ownership table, the original query is dynamically rewritten into a guarded query with an additional owner predicate, as shown below:

\begin{lstlisting}[basicstyle=\ttfamily\footnotesize, escapechar=|, frame=single, backgroundcolor=\color{gray!5}]
|\textcolor{teal}{// Original SQL}|
SELECT * FROM sys_notice WHERE notice_id = ?

|\textcolor{teal}{// Rewritten Guarded SQL}|
SELECT * FROM sys_notice WHERE notice_id = ?
|\textcolor{blue}{AND create\_by = currentUser}|
\end{lstlisting}

The injected value is derived from the server-side identity record rather than from attacker-controlled request parameters. Therefore, changing an object identifier in the HTTP request is insufficient to bypass the guard.

To support external applications, we implemented a Java agent. The agent instruments JDBC \texttt{prepareStatement} calls in common connection implementations, including MySQL JDBC and Druid. Before a candidate SQL statement is prepared, the agent checks whether the SQL operation and table match the configured metadata. If so, it sends the SQL and identity record to the guard server. The guard server returns one of three access-control decisions: \texttt{ALLOW}, \texttt{REWRITE}, or \texttt{DENY}. An \texttt{ALLOW} decision indicates that the target object belongs to the current user or is a public resource, requiring no modification. A \texttt{REWRITE} decision indicates that the table is protected and requires AST-level injection of the current user's identity constraint. A \texttt{DENY} decision is issued when the required identity information is unavailable due to propagation failure, enforcing a fail-closed behavior that aborts the operation before database execution.

To bridge the semantic gap between application-level identities and JDBC executions in real-world systems, we implemented framework-specific identity extraction hooks for representative open-source enterprise systems, including RuoYi, BootDo, and XXL-Job \citep{ruoyiProject,bootdoProject,xxljobProject}. By intercepting Shiro or custom authentication filters, the extracted \texttt{userId}, \texttt{username}, and roles are reliably bound to the identity record without modifying legacy business code.

For performance, the agent implements a rewrite-decision cache. The cache key contains the application name, identity record, and a fingerprint of the normalized SQL. This optimization ensures that repeated executions of high-frequency hot templates avoid redundant parsing and inter-process communication overhead.

\section{Evaluation}
\label{sec:evaluation}

\begin{figure*}[t]
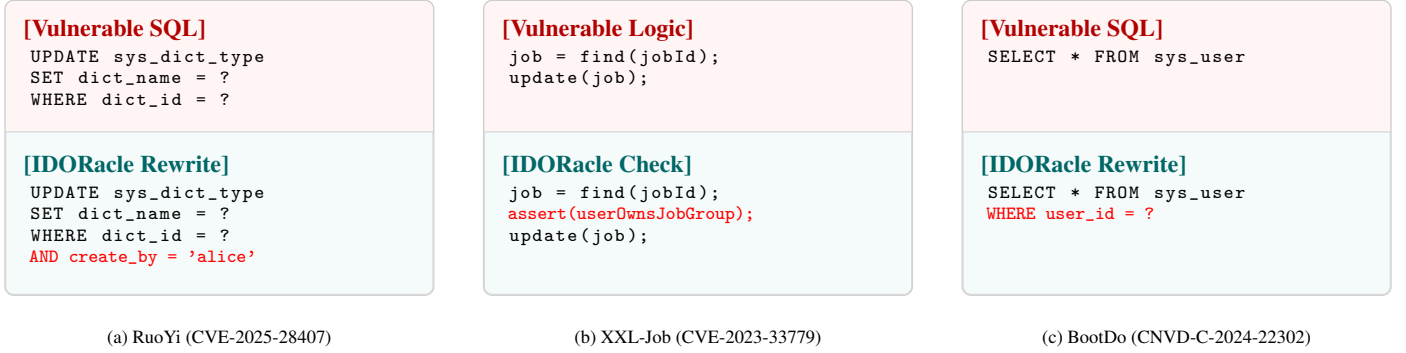

    \centering
    
    \begin{subfigure}[t]{0.31\textwidth}
        \begin{tcolorbox}[
            bicolor,
            colframe=gray!40,
            colback=red!4,
            colbacklower=teal!4,
            boxrule=0.5pt, arc=3pt,
            top=4pt, bottom=4pt, left=4pt, right=4pt,
            segmentation style={solid, draw=gray!30, line width=0.5pt}
        ]
            \begin{minipage}[t][1.2cm][t]{\linewidth}
                \textcolor{red!70!black}{\textbf{\small [Vulnerable SQL]}}
                \vspace{0.2em}
                \begin{lstlisting}[basicstyle=\ttfamily\scriptsize, frame=none, xleftmargin=2pt, aboveskip=0pt, belowskip=0pt]
UPDATE sys_dict_type 
SET dict_name = ? 
WHERE dict_id = ?
                \end{lstlisting}
            \end{minipage}
            
            \tcblower
            
            \begin{minipage}[t][1.6cm][t]{\linewidth}
                \textcolor{teal!80!black}{\textbf{\small [IDORacle Rewrite]}}
                \vspace{0.2em}
                \begin{lstlisting}[basicstyle=\ttfamily\scriptsize, frame=none, escapechar=|, xleftmargin=2pt, aboveskip=0pt, belowskip=0pt]
UPDATE sys_dict_type 
SET dict_name = ? 
WHERE dict_id = ? 
|\textcolor{red}{\textbf{AND create\_by = 'alice'}}|
                \end{lstlisting}
            \end{minipage}
        \end{tcolorbox}
        \caption{RuoYi (CVE-2025-28407)}
        \label{fig:cve_ruoyi}
    \end{subfigure}
    \hfill
    \begin{subfigure}[t]{0.31\textwidth}
        \begin{tcolorbox}[
            bicolor, colframe=gray!40, colback=red!4, colbacklower=teal!4, 
            boxrule=0.5pt, arc=3pt, top=4pt, bottom=4pt, left=4pt, right=4pt,
            segmentation style={solid, draw=gray!30, line width=0.5pt}
        ]
            \begin{minipage}[t][1.2cm][t]{\linewidth}
                \textcolor{red!70!black}{\textbf{\small [Vulnerable Logic]}}
                \vspace{0.2em}
                \begin{lstlisting}[basicstyle=\ttfamily\scriptsize, frame=none, xleftmargin=2pt, aboveskip=0pt, belowskip=0pt]
job = find(jobId);
update(job);
                \end{lstlisting}
            \end{minipage}
            
            \tcblower
            
            \begin{minipage}[t][1.6cm][t]{\linewidth}
                \textcolor{teal!80!black}{\textbf{\small [IDORacle Check]}}
                \vspace{0.2em}
                \begin{lstlisting}[basicstyle=\ttfamily\scriptsize, frame=none, escapechar=|, xleftmargin=2pt, aboveskip=0pt, belowskip=0pt]
job = find(jobId);
|\textcolor{red}{\textbf{assert(userOwnsJobGroup);}}|
update(job);
                \end{lstlisting}
            \end{minipage}
        \end{tcolorbox}
        \caption{XXL-Job (CVE-2023-33779)}
        \label{fig:cve_xxljob}
    \end{subfigure}
    \hfill
    \begin{subfigure}[t]{0.31\textwidth}
        \begin{tcolorbox}[
            bicolor, colframe=gray!40, colback=red!4, colbacklower=teal!4, 
            boxrule=0.5pt, arc=3pt, top=4pt, bottom=4pt, left=4pt, right=4pt,
            segmentation style={solid, draw=gray!30, line width=0.5pt}
        ]
            \begin{minipage}[t][1.2cm][t]{\linewidth}
                \textcolor{red!70!black}{\textbf{\small [Vulnerable SQL]}}
                \vspace{0.2em}
                \begin{lstlisting}[basicstyle=\ttfamily\scriptsize, frame=none, xleftmargin=2pt, aboveskip=0pt, belowskip=0pt]
SELECT * FROM sys_user
                \end{lstlisting}
            \end{minipage}
            
            \tcblower
            
            \begin{minipage}[t][1.6cm][t]{\linewidth}
                \textcolor{teal!80!black}{\textbf{\small [IDORacle Rewrite]}}
                \vspace{0.2em}
                \begin{lstlisting}[basicstyle=\ttfamily\scriptsize, frame=none, escapechar=|, xleftmargin=2pt, aboveskip=0pt, belowskip=0pt]
SELECT * FROM sys_user 
|\textcolor{red}{\textbf{WHERE user\_id = ?}}|
                \end{lstlisting}
            \end{minipage}
        \end{tcolorbox}
        \caption{BootDo (CNVD-C-2024-22302)}
        \label{fig:cve_bootdo}
    \end{subfigure}

    \caption{Comparison of vulnerable database operations and IDORacle's runtime mediation outcomes across three real-world vulnerabilities.}
    \label{fig:cve_rewrites}
\end{figure*}

\begin{table*}[t]
\centering
\caption{Evaluation resources and ownership assignments used in the real-world case study.}
\label{tab:realworld_dataset}
\small
\setlength{\tabcolsep}{10pt}
\begin{tabular}{lcclcc}
\toprule
\textbf{Resource Type} & \textbf{Resource ID} & \textbf{Resource Name} & \textbf{Ownership Attribute} & \textbf{Owner} & \textbf{Related CVE} \\
\midrule
Dictionary Type & 9001 & Alice Dictionary Type & \texttt{create\_by=alice} & Alice & CVE-2025-28407 \\
                & 9002 & Bob Dictionary Type   & \texttt{create\_by=bob}   & Bob   & CVE-2025-28407 \\
\addlinespace[0.6em]
Notice          & 9001 & Alice Notice          & \texttt{create\_by=alice} & Alice & CVE-2025-28412 \\
                & 9002 & Bob Notice            & \texttt{create\_by=bob}   & Bob   & CVE-2025-28412 \\
\addlinespace[0.6em]
Scheduled Job   & 9001 & Alice Job             & \texttt{create\_by=alice} & Alice & CVE-2025-28402 \\
                & 9002 & Bob Job               & \texttt{create\_by=bob}   & Bob   & CVE-2025-28402 \\

\bottomrule
\end{tabular}
\end{table*}

\begin{figure*}[t]
  \centering
  \includegraphics[width=\textwidth]{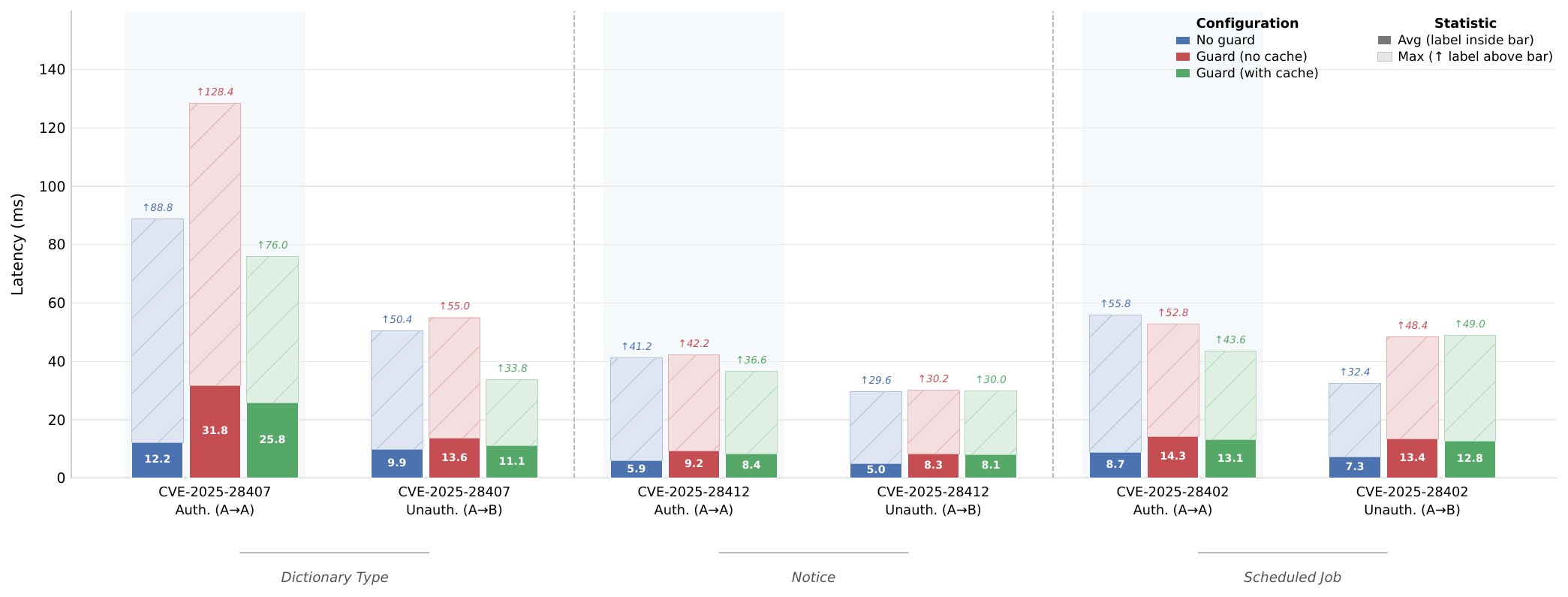}
  \caption{End to end request latency per execution mode across three RuoYi IDOR vulnerabilities reporting mean and maximum performance values.}
  \label{fig:real_world_latency}
\end{figure*}

We evaluate IDORacle to assess its security efficacy, practical deployability, and runtime efficiency through a series of empirical experiments. The evaluation addresses the following four distinct research questions:

\begin{itemize}
  \item \textbf{RQ1 (Security Effectiveness):} Does IDORacle prevent horizontal object reference violations across diverse Java SQL scenarios without introducing false positives? (Section~\ref{subsec:security_effectiveness})
  \item \textbf{RQ2 (Practical Deployability):} Can the framework integrate with real world legacy Java web applications to mitigate documented CVEs? (Section~\ref{subsec:practical_deployability})
  \item \textbf{RQ3 (Micro Level Performance):} What is the latency overhead introduced by the SQL sink reference monitor and how effective is the template caching mechanism? (Section~\ref{subsec:performance})
  \item \textbf{RQ4 (End to End Impact):} How does IDORacle affect the end to end response latency of web endpoints under realistic operational conditions? (Section~\ref{subsec:real_world_case_study})
\end{itemize}

\subsection{Experimental Setup}
\label{subsec:evaluation_setup}

The prototype evaluation is conducted on a dedicated testbed utilizing the Spring Boot framework version 2.7 with MyBatis 3.5 and a MySQL 8.0 database management system. The underlying storage layer leverages HikariCP for connection pooling under a shared credential architecture. The hardware environment comprises an Intel Core i7 processor operating at 3.6 GHz paired with 16 GB of memory running an Ubuntu 20.04 long term support operating system. We evaluate and compare three operational configurations. The \textbf{NO\_GUARD} configuration executes queries in their original state to establish the baseline database behavior. The \textbf{GUARD\_NO\_CACHE} configuration enforces the entire reference monitor validation flow including parsing and guard matching for every single database statement. The \textbf{GUARD\_WITH\_CACHE} configuration activates the dual fingerprint template plan cache to eliminate redundant analytical overhead.

\subsection{RQ1: Security Effectiveness}
\label{subsec:security_effectiveness}

To evaluate the fundamental security efficacy of IDORacle, we execute paired benign and adversarial execution traces across diverse object level authorization configurations. The experimental data reveals a structural vulnerability pattern in legacy applications where database mappers execute queries relying entirely on request supplied resource identifiers. We classify the evaluated data access paths into two structural paradigms consisting of direct ownership validation and complex derived relational validation.

For the direct ownership paradigm represented by tables such as sys dict type and sys notice, the primary threat stems from simple parameter substitution. The baseline data demonstrates that an adversary can retrieve data belonging to other tenants without restriction. Under the IDORacle configuration, the framework dynamically intercepts the query at the database boundary and appends an explicit user identifier predicate to the abstract syntax tree structure. When Alice attempts an unauthorized access targeting Bob, the injected constraint forces the database engine to evaluate an empty result set. The application subsequently translates this empty payload into a safe response class such as an HTTP 404 error or an empty data list, successfully neutralizing the horizontal escalation attempt.

For the derived relational validation paradigm where authorization evidence resides in secondary tables, the security risks expand significantly due to the absence of explicit owner columns in the primary target sink. IDORacle addresses this challenge by generating complex subqueries or executing the probe then allow fallback mechanism. The results confirm a one hundred percent attack prevention rate across all synthesized scenarios with zero false positives. Legitimate operations executed by true resource owners consistently pass validation without semantic distortion or data reduction. However, the evaluation also indicates that highly convoluted analytical structures containing deeply nested aggregations restrict safe rewrite options, requiring the mediator to degrade gracefully to standalone validation probes which demands optimization to sustain operational throughput.

\subsection{RQ2: Practical Deployability}
\label{subsec:practical_deployability}

We assess the practical deployability of the framework by installing the automated Java agent into three widely adopted open source enterprise platforms consisting of RuoYi, BootDo, and XXL Job. The underlying operational problem in these legacy architectures is the complete disconnection between upstream web authentication and downstream persistence layers, as data access calls operate entirely via anonymized connection pool flows. We classify the evaluation by framework integration types spanning Apache Shiro security configurations and custom filter structures.

The Java agent achieves transparent enforcement across all target applications without requiring any manual source code modifications to existing business services or mapper XML structures. We validate this capability against four distinct real world vulnerabilities comprising CVE 2025 28407, CVE 2025 28412, CVE 2025 28402, and CVE 2025 28413. The agent successfully hooks the lower level JDBC prepareStatement interface and intercepts every candidate query before database interaction. By extracting identity contexts from Shiro session attributes and mapping them to downstream trace identifiers, the framework establishes a reliable identity propagation bridge across asynchronous execution boundaries.

The deployment evaluation demonstrates that IDORacle functions as a non intrusive safety net for legacy infrastructure where retrofitting security annotations directly into scattered service modules is technically impractical. Nevertheless, the empirical findings also underscore an inherent architectural constraint. The deployment success is strictly contingent upon the integrity of the primary upstream authentication tier. Should the initial JWT or session validation component be entirely compromised by an attacker, the metadata bound to the server side trace identity becomes invalid, which validates IDORacle as a powerful defense in depth mechanism rather than a total replacement for fundamental entry control.

\subsection{RQ3: Micro Level Performance}
\label{subsec:performance}

To evaluate the micro level computational cost introduced by the database reference monitor, we subject the oracle engine to high concurrency stress tests under varying cache states. The baseline NO GUARD processing latency remains below 0.001 ms per query instance. In the worst case GUARD NO CACHE scenario, where the engine is forced to perform complete lexical parsing, table classification, context extraction, and abstract syntax tree rewrite generation for every execution instance, the mean latency reaches 0.17 ms. This performance penalty introduces a discernible bottleneck that scales linearly with query frequency inside batch application loops.

The activation of the redundancy aware template plan cache in the GUARD WITH CACHE configuration significantly mitigates this performance limitation. By utilizing normalized SQL template fingerprints and identity context records to look up precompiled mediation strategies, the mean processing cost drops sharply to 0.017 ms, representing an average performance optimization of ninety percent. This reduction demonstrates that the system effectively separates the expensive initial structural analysis from the lightweight per instance parameter binding work.

However, a granular evaluation of the cache behavior reveals a specific operational variance depending on the type of persistence framework used by the host application. While static mapping tools such as MyBatis maintain stable templates that yield near perfect cache hit ratios, fully dynamic Object Relational Mapping implementations such as Hibernate or JPA often generate highly fragmented and volatile raw SQL texts at runtime. In such environments, the template plan fingerprint uniqueness increases, causing a reduction in the overall cache hit rate and driving performance down toward the uncached baseline. This finding highlights that lifting the interception boundary to higher level abstraction APIs constitutes an essential requirement for highly dynamic data access architectures.

\subsection{RQ4: End to End Impact}
\label{subsec:real_world_case_study}

We examine the macroscopic impact of the defense framework by tracking the end to end response latency of individual web endpoints across the four reproduced enterprise vulnerabilities. As illustrated in Figure 3, the evaluation encompasses four specific indicators comprising the average total latency, the average request latency, the maximum request ceiling, and the average minimum request baseline. The experimental results reveal a stark contrast in the baseline performance characteristics across different business domains. For example, Dictionary Type operations under CVE 2025 28407 exhibit a higher inherent mean latency of approximately 30 ms with maximum spikes exceeding 120 ms, whereas Notice management endpoints under CVE 2025 28412 operate within a tighter distribution averaging around 10 ms due to simpler configuration lookups and fewer internal join operations.

When analyzing the execution behavior, we classify the requests into authorized self access paths and unauthorized horizontal privilege escalation paths. For authorized operations, the reference monitor introduces a marginal latency increase across all endpoints due to the mandatory interception and context extraction steps. The caching mechanism successfully absorbs a major fraction of this cost, as evidenced by the average minimum request indicator where cached requests consistently run several milliseconds faster than their uncached counterparts.

A highly compelling and non intuitive finding emerges when analyzing the unauthorized horizontal escalation paths where Alice targets resources owned by Bob. Across all evaluated CVE scenarios, the end to end latency for blocked or rewritten adversarial requests is consistently lower than the latency recorded for the corresponding authorized paths. This phenomenon is directly attributable to the structural effects of the AST rewrite on the database engine execution plan. Because the injected user identity predicate restricts row visibility to Alice, the query yields an empty result set early in the database processing pipeline. This early termination short circuits heavy internal data serialization, memory allocation, and large network payload transfers back to the application server. Consequently, the defense mechanism turns a security enforcement boundary into an accidental performance optimization for malicious traffic, ensuring that the average end to end response latency remains safely below 35 ms even under intensive automated attack campaigns.

\section{Discussion}
\label{sec:discussion}

\subsection{Scope of SQL-Sink Enforcement}

IDORacle is designed to prevent horizontal object-level authorization
violations whose security effect is ultimately realized through Java-SQL
database operations. Its protection boundary is intentionally narrower than
the full space of access-control failures. The system is effective when an
attacker-controlled object identifier reaches a protected SQL sink and when
the relevant authorization relation can be represented by table metadata,
ownership predicates, or column policies.

Several limitations inherently bound this architectural scope. First,
privilege escalation requests executing through third-party opaque APIs,
rather than downstream observable SQL sinks, evade interception. Without
local protected SQL operations, IDORacle lacks the runtime AST manipulation
leverage necessary to enforce constraints. Second, accesses that bypass
instrumented JDBC drivers---such as complex encapsulated stored procedures
without explicit resource semantics---require manual adapter interventions.
Third, if access-control decisions rely exclusively on volatile transient
business states unmapped to database schemas or the identity record, a
purely SQL-layer reference monitor cannot independently infer such intent.
Fourth, fully automated ORM frameworks such as Hibernate or JPA often
generate highly dynamic and fragmented SQL strings at runtime; applying
IDORacle directly at the raw JDBC layer in such environments may reduce
template cache hit rates, potentially approaching the \texttt{GUARD\_NO\_CACHE}
performance baseline ($\sim$0.17\,ms). A viable mitigation is to lift the
interception point to the HQL or Criteria API abstraction layer, where
query structure remains stable and the dual-fingerprint mechanism retains
its efficiency---though this requires framework-specific adapters beyond the
current prototype.

A broader avenue for future work is the integration of IDORacle with
static analysis or authorization-inference tools such as BolaRay or
MOCGuard~\citep{huang2024bolaray,liu2025mocguard}. Automatically generating
baseline policy templates from inferred ownership relations would reduce
the manual configuration overhead for large-scale legacy deployments,
allowing IDORacle to serve as the runtime enforcement engine for
statically derived policies.

\subsection{Layer Placement and Deployment Trade-offs}

IDORacle does not require developers to rewrite MVC controllers, service
methods, or mapper definitions. This low-intrusion deployment model is
important for legacy Java systems, where authorization logic is often
scattered. By enforcing object-level constraints at the DAO/JDBC boundary,
IDORacle provides a resilient safety net even when service-layer checks are
absent or incomplete.

However, delaying enforcement to the JDBC layer introduces specific
deployment trade-offs. When a malicious request could have been denied at
the routing layer via strict RBAC annotations, intercepting it at the SQL
sink means the application has already expended CPU cycles on routing,
parameter parsing, and intermediate business logic. While this late-enforcement
model ensures the database state remains secure, it incurs a marginal
computational cost prior to rejection. Nonetheless, as the micro-benchmark
and caching results demonstrate, this overhead is effectively contained.
IDORacle prioritizes a deterministic, unified enforcement point at the
data-access boundary, complementing rather than supplanting existing early-stage authorization mechanisms.

\section{Related Work}
\label{sec:related-work}

\subsection{Web Logic Vulnerabilities}

Access-control flaws are widely recognized as semantic web logic vulnerabilities, distinct from syntactic input-validation errors. Early systems such as NoTamper, MACE, and AuthScope reason about hidden parameters, user sessions, and protocol fields to identify parameter tampering and horizontal privilege escalation \citep{bisht2010notamper,monshizadeh2014mace,zuo2017authscope}. Complementary static and black-box studies further demonstrate that authorization defects depend on application-specific workflows, state transitions, and business invariants \citep{sun2011staticac,felmetsger2010logic,pellegrino2014logicflaws,wang2026anota}. These analyses establish that the legitimacy of a given request parameter is determined by the authenticated subject, the referenced object, and the backend operation jointly---a perspective that motivates treating IDOR prevention as a subject--object--operation consistency problem.

Building on this foundation, recent testing systems improve logic-vulnerability discovery through stateful crawling, directed fuzzing, multi-user differential testing, and runtime validation. Atropos, EvoCrawl, and Predator emphasize that realistic vulnerability discovery requires valid sessions, reachable program states, and semantically meaningful inputs \citep{guler2024atropos,guo2025evocrawl,wang2025predator}. A2CT and DRacv further address automated testing and repair for function- and role-based access-control flaws \citep{schlaubitz2025a2ct,xu2025dracv}. These techniques are complementary to IDORacle: they expose access-control defects during testing and analysis phases, whereas this work focuses on enforcing object-level authorization for deployed Java-SQL applications where vulnerable code paths may persist in production.

\subsection{BOLA Detection and Ownership Analysis}

IDOR is characterized as a concrete instance of Broken Object-Level Authorization (BOLA), where an authenticated user manipulates an object reference to access data or trigger state transitions outside the authorized scope \citep{owasp2023bola,cwe639,portswiggerIDOR}. Unlike Broken Function-Level Authorization, the endpoint itself may be legitimately callable; the violation arises when the concrete object selected by the request is not owned by, visible to, or administrable by the current subject \citep{owasp2023bfla,owasp2023bopla}. Effective IDOR reasoning therefore requires connecting the authenticated principal, attacker-controlled object identifiers, and the backend operation that dereferences them.

Recent BOLA-oriented systems provide increasingly precise evidence for missing object-level checks. BolaRay infers object-level authorization models in database-backed applications, while MOCGuard targets missing-owner-check vulnerabilities in Java web applications \citep{huang2024bolaray,liu2025mocguard}. BACScan and BACFuzz demonstrate that HTTP response observation alone may be insufficient, and improve detection through multi-user differential analysis, runtime information, and SQL-level evidence \citep{liu2025bacscan,dharmaadi2025bacfuzz}. UABScan further reveals inconsistent authentication and authorization assumptions across Java web layers \citep{zhang2025uabscan}. Empirical and zero-trust-oriented studies reinforce the need to reason about BOLA beyond simple identifier guessing \citep{wu2025bolaz,kaur2026bolawild}. Collectively, these works advance the state of vulnerability discovery. IDORacle is positioned differently: it shifts the objective from detecting whether a vulnerable endpoint exists to preventing unauthorized object operations at the SQL execution boundary, providing a runtime safety net for applications where detection-phase remediation is incomplete.

\subsection{SQL-Level Policy Enforcement}

Database and middleware research has long explored fine-grained policy enforcement through query rewriting, authorization views, predicated grants, row-level security, and policy-aware adapters. Classical approaches encode fine-grained authorization decisions as SQL predicates inside query processing \citep{rizvi2004nontruman,chaudhuri2007predicatedgrants}, and modern database systems provide row-level-security mechanisms for policy-controlled row visibility \citep{postgresqlRLS,sqlserverRLS}. Security systems such as Qapla, Authorized Updates, and Blockaid further demonstrate that application-layer data access can be protected by mediating SQL outside ordinary business logic \citep{mehta2017qapla,eykholt2017authorizedupdates,zhang2022blockaid}. Related middleware and policy-extraction studies address scalable database access control, query-control interfaces, and automated recovery of application access-control policies \citep{pappachan2020sieve,bogaerts2021sequoia,shay2018querycontrol,zhang2023beyondpolicy,zhang2024ote}.

While these systems establish the feasibility of SQL-level mediation, typical Java web deployments introduce an additional semantic gap not addressed by prior work. JDBC connection pools execute SQL under shared database credentials, rendering the underlying DBMS unable to observe end-user identity. Authenticated user context, maintained in framework security components such as filters, controllers, or service methods, is therefore invisible at the database boundary. Moreover, IDOR policies span direct owner columns, join-derived ownership, group or tenant membership, role-sensitive transitions, and sensitive-column constraints---a breadth not captured by generic query-rewriting or row-level-security primitives. This gap motivates a Java-SQL-specific design that binds server-side identity context to SQL-template mediation, as realized by IDORacle.

\subsection{Java-SQL Access Control}

RBAC and ABAC remain important frameworks for expressing roles, attributes, and coarse-grained authorization policies \citep{sandhu1996rbac,hu2014abac}. In legacy Java enterprise systems, however, authorization logic is distributed across filters, annotations, controllers, service methods, MyBatis mappers, scheduled jobs, and JDBC calls. The layer that holds the authenticated subject identity is typically separated from the layer that executes the concrete SQL statement against the database. This separation creates a last-mile enforcement gap that upstream framework mechanisms do not directly address.

IDORacle targets this gap by placing a reference monitor at the MyBatis/JDBC boundary as a complementary enforcement point, rather than a replacement for upstream authentication or role authorization. Its template-guided design amortizes expensive parsing, normalization, table classification, and guard binding at the SQL-template level. Per-instance mediation then performs context binding, resource-key extraction, proof lookup, and guard-specific validation with low overhead. This placement preserves existing business code while enforcing object-level constraints immediately before protected SQL statements reach the database, making it well-suited for legacy Java-SQL applications where retrofitting authorization into service methods is impractical.

\section{Conclusion}
\label{sec:conclusion}

This paper presented IDORacle, a template-guided runtime prevention framework for
Java-SQL IDOR vulnerabilities. IDORacle bridges the semantic gap between
application-level identity and SQL-level object operations by propagating trusted
subject context, compiling reusable SQL-template mediation plans, and enforcing
ownership, membership, resolver-based, role/state, and sensitive-column guards
before protected statements reach the database. The evaluation shows that IDORacle
prevents the tested horizontal object-reference violations while preserving benign
accesses, with a worst-case guard latency of 0.17\,ms and a 90.1\% overhead
reduction under rewrite-decision caching. These results demonstrate that SQL-sink
mediation can provide a practical last-mile safety net for legacy Java
database-backed applications, complementing existing controller-, service-, and
external-API authorization mechanisms.

\section*{Declaration of competing interest}
The authors declare that they have no known competing financial interests or
personal relationships that could have appeared to influence the work reported
in this paper.

\section*{Data availability}
Data will be made available on request.

\section*{Funding}
This research did not receive any specific grant from funding agencies in the
public, commercial, or not-for-profit sectors.

\bibliographystyle{elsarticle-harv}
\bibliography{bibliography}

\end{document}